\documentstyle[12pt]{article}
\newlength{\extraspace}
\newlength{\extraspaces}
\newcommand{\be}{\begin{equation}
\addtolength{\abovedisplayskip}{\extraspaces}
\addtolength{\belowdisplayskip}{\extraspaces}
\addtolength{\abovedisplayshortskip}{\extraspace}
\addtolength{\belowdisplayshortskip}{\extraspace}}
\newcommand{\ee}{\end{equation}}

\newcommand{\ba}{\begin{eqnarray}
\addtolength{\abovedisplayskip}{\extraspaces}
\addtolength{\belowdisplayskip}{\extraspaces}
\addtolength{\abovedisplayshortskip}{\extraspace}
\addtolength{\belowdisplayshortskip}{\extraspace}}
\newcommand{\ea}{\end{eqnarray}}

\newcommand{\nonu}{\nonumber \\[.5mm]}
\newcommand{\A}{&\!\!\!}

\newcommand{\il}{\lambda_{{}_{{}_{\!\!\!\!\scriptstyle{i}}}}}

\newcommand{\bl}{\lambda_{{}_{{}_{\!\!\!\!\scriptstyle{0}}}}}

\newcommand{\blz}{\lambda_{{}_{{}_{\!\!\!\!\scriptstyle{0}}}}}

\newcommand{\al}{\lambda_{{}_{{}_{\!\!\!\!\scriptstyle{a}}}}}

\newcommand{\newsection}[1]{
\vspace{7mm} \pagebreak[3] \addtocounter{section}{1}
\setcounter{subsection}{0} \setcounter{footnote}{0}
\begin{center}
{\large {\bf \thesection. #1}}
\end{center}
\nopagebreak
\medskip
\nopagebreak \hspace{3mm}}

\begin{document}
\begin{center}
{{\bf Vacuum Non Singular Black Hole in Tetrad Theory of
Gravitation}}
\end{center}
\centerline{ Gamal G.L. Nashed}

\bigskip

\centerline{{\it Mathematics Department, Faculty of Science, Ain
Shams University, Cairo, Egypt }}

\bigskip
 \centerline{ e-mail:nasshed@asunet.shams.eun.eg}

\hspace{2cm}
\\
\\
\\
\\
\\
\\
\\
\\

The field equations of a special class of tetrad theory of
gravitation have been applied to tetrad space having  three
unknown functions of radial coordinate. The spherically symmetric
vacuum stress-energy momentum tensor with one assumption
concerning  its specific form generates two non-trivial different
exact analytic solutions for these field equations. For large $r$,
the exact analytic solutions coincide with the Schwarzschild
solution, while for small $r$, they  behave in a manner similar to
the de Sitter solution and describe a spherically symmetric black
 hole singularity free everywhere. The solutions obtained give rise
to two different tetrad structures, but having the same metric,
i.e., a static spherically symmetric nonsingular black hole
metric. We then, calculated the energy associated with these two
exact analytic solutions using the superpotential method. We find
that unless the time-space components of the tetrad go to zero
faster than $ \displaystyle{1 \over \sqrt{r}}$ at infinity, the
two solutions give different results. This fact implies that the
time-space components of the tetrad must vanish faster than
$\displaystyle{1 \over \sqrt{r}}$ at infinity.

\newpage
\begin{center}
\newsection{\bf Introduction}
\end{center}

In the tetradic theories of gravitation the gravitational field is
described by a tetrad field. In the past, several authors  have
considered these theories to solve some problems of general
relativity or to extend it to include  other fields. While
searching for a unification of gravitation and electromagnetism
Einstein considered a tetrad theory  \cite{Ei}, but soon he
rejected it because there was no Schwarzschild solution in his
field equation.

 M\o ller  has shown that it is impossible to find a satisfactory
 expression for the energy-momentum complex in the framework of Riemannian geometry
  \cite{Mo}. In a series of papers, \cite{Mo,Mo3,Mo4} M\o ller was able to obtain
  a general expression for a satisfactory energy-momentum complex in the absolute
  parallelism space. The Lagrangian formulation of this tetrad
theory of gravitation was first given by Pellegrini and Plebanski
\cite{PP}. In these attempts the admissible Lagrangians were
limited by the assumption that the equations determining the
metric tensor should coincide with the Einstein equation.  M\o
ller \cite{Mo1} abandoned this assumption and suggested to look
for a wider class of Lagrangians, allowing for possible deviation
from the Einstein equation in the case of strong gravitational
fields. S$\acute{a}$ez \cite{Se} generalized M\o ller theory into
a scalar tetrad theory of gravitation.  Meyer \cite{Me} showed
that M\o ller theory is a special case of Poincar$\acute{e}$ gauge
theory \cite{HS,HNV}.

Hayashi and  Nakano \cite{HN} formulated the tetrad theory of
gravitation as the gauge theory of space-time translation group.
Hayashi and  Shirafuji \cite{HS1} studied the geometrical and
observational basis of the tetrad  theory, assuming that the
Lagrangian  be given by a quadratic form of torsion tensor
assuming the Lagrangian to be invariant under parity operation
involving three unknown parameters to be fixed by experiments,
besides a cosmological term. Two of the three parameters were
determined by comparing with solar-system experiments, \cite{HS1}
while only an upper bound has been estimated for the third one
\cite{HS1,MN1}.

The numerical values of the two parameters found were very small
consistent with being equal to zero. If these two parameters are
equal to zero exactly, the theory reduces to the one proposed by
Hayashi and  Nakano \cite{HN} and M\o ller, \cite{Mo1} which we
shall here refer to as the HNM theory for short. This theory
differs from general relativity only when the torsion tensor has
nonvanishing axial-vector part. It was also  shown \cite{HS1} that
the Birkhoff theorem can be extended to the HNM theory. Namely,
for spherically symmetric case in vacuum, which is not necessarily
time independent, the axial-vector part of the torsion tensor
should vanish due to the antisymmetric
 part of the field equation, and therefore, with the help of the Birkhoff
 theorem \cite{Br} of general relativity we see that the spacetime metric
is the Schwarzschild.

Mikhail et al. \cite{MWHL} derived the superpotential of the
energy-momentum complex in the HNM theory and applied it to two
spherically symmetric solutions. It was found that in one of the
two solutions the gravitational mass does not coincide with the
calculated energy. This result was extended to a wider class of
solutions with spherical symmetry \cite{SNH}. An explicit
expression was given for all the stationary, asymptotically flat
solutions with ${\it spherical}$ ${\it symmetry}$, which were then
classified according to the asymptotic behavior of the components
of $({\al}_{\,{}^{{}^{{}^{\scriptstyle{0}}}}})$ and
$({\bl}_{\,{}^{{}^{{}^{\scriptstyle{\alpha}}}}})$. It was found
that the equality of the gravitational and inertial masses holds
only when $({\al}_{\,{}^{{}^{{}^{\scriptstyle{0}}}}})$ and
$({\bl}_{\,{}^{{}^{{}^{\scriptstyle{\alpha}}}}})$ tend to zero
faster
 than $1/{\sqrt{r}}$.
\newpage

Dymnikova \cite{Di} derived a static spherically symmetric
nonsingular black hole solution in orthodox general relativity
assuming a specific form of the stress-energy momentum tensor.
This solution practically coincides with the Schwarzschild
solution for large $r$,  for small $r$ behaves like the de Sitter
solution and describe a spherically symmetric black hole
singularity free everywhere \cite{Di}.

The general form of the tetrad, $\il^\mu$, having spherical
symmetry was given by  Robertson  \cite{Ro}. In the Cartesian form
it can be written as \footnote{In this paper Latin indices
$(i,j,...)$ represent the vector number, and Greek indices
$(\mu,\nu,...)$ represent the vector components. All indices run
from 0 to 3. The spatial part of Latin indices are denoted by
$(a,b,...)$, while that of greek indices by $(\alpha, \beta,...).$
 In the present convention, latin indices are never raised. The tetrad
$\il^\mu$ is related to the parallel vector fields ${b_i}^\mu$ of
\cite{HS1} by $\blz^\mu=i{b_{0}}^\mu$ and $\al^\mu={b_a}^\mu$.}

\ba \bl^0 \A= \A iA, \quad \al^0 = C x^a, \quad \bl^\alpha = iD
x^\alpha \nonu
\al^\alpha \A= \A \delta_a^\alpha B + F x^a x^\alpha + \epsilon_{a
\alpha \beta} S x^\beta, \ea where {\it A}, {\it C}, {\it D}, {\it
B}, {\it F}, and {\it S} are functions of ${\it t}$ and
$r=(x^\alpha x^\alpha)^{1/2}$, and the zeroth vector $\bl^\mu$ has
the factor $i=\sqrt{-1}$ to preserve Lorentz signature. We
consider an asymptotically flat space-time in this paper, and
impose the boundary condition that for $r \rightarrow \infty$
 the tetrad (1) approaches the tetrad of Minkowski space-time,
$\left(\il^\mu \right)= {\rm diag}(i,{\delta_a}^{\alpha})$.
Robertson has shown that:\vspace{.2cm}\\ 1-Improper rotation are
admitted if and only if $S=0$.\vspace{.3cm}\\ 2- The functions C
and F can be eliminated by a mere coordinate transformations,
i.e., by making use of freedom to redefine $t$ and $r$, leaving
the tetrad (1) having three unknown functions in the Cartesian
coordinate, which will be used in section 3 for calculations of
HNM  field equations but in the spherical polar coordinate.

It is the aim of the present work to find the, asymptotically flat
solutions with spherical symmetry which is different from the
Schwarzschild solution in the HNM theory. Assuming the same form
of the stress-energy momentum tensor as  given by Dymnikova
\cite{Di}, we obtain two different exact analytic solutions of the
HNM theory. We then, calculate the energy of those solutions using
the superpotential of Mikhail et
 al. \cite{MWHL}.

In section 2 we briefly review the tetrad theory of gravitation.
In section 3 we first apply the tetrad (1) with three unknown
functions of the radial coordinate in spherical polar coordinates
to the field equations of HNM theory, then we derived the
corresponding partial differential equations. Assuming some
conditions on these partial differential equations we obtained two
different exact asymptotically flat solutions with spherical
symmetry. In section 4 the energy of the gravitating source is
calculated by the superpotential method. The final section is
devoted to the main results and discussion.

Computer algebra system Maple V Release 4 is used in some
calculations.

\newpage
\newsection{The tetrad theory of gravitation}

In this paper we follow M\o ller's construction of the tetrad
theory of gravitation based on the Weitzenb{\rm $\ddot{o}$}ck
space-time. In this theory the field variables are the 16 tetrad
components $\il^\mu$, from which the metric is derived by \be
g^{\mu \nu} \stackrel{\rm def.}{=} \il^\mu \il^\nu. \ee The
Lagrangian ${\it L}$ is an invariant constructed from $\gamma_{\mu
\nu \rho}$ and $g^{\mu \nu}$, where $\gamma_{\mu \nu \rho}$ is the
contorsion tensor given by \be \gamma_{\mu \nu \rho} \stackrel{\rm
def.}{=} {\il}_{\,{}^{{}^{{}^{\scriptstyle{\mu}}}}}
{\il}_{\,{}^{{}^{{}^{\scriptstyle{\nu;\rho}}}}}, \ee where the
semicolon denotes covariant differentiation with respect to
Christoffel symbols. The most general Lagrangian density invariant
under parity operation is given by the form \be {\cal L}
\stackrel{\rm def.}{=} (-g)^{1/2} \left( \alpha_1 \Phi^\mu
\Phi_\mu+ \alpha_2 \gamma^{\mu \nu \rho} \gamma_{\mu \nu \rho}+
\alpha_3 \gamma^{\mu \nu \rho} \gamma_{\rho \nu \mu} \right), \ee
where \be g \stackrel{\rm def.}{=} {\rm det}(g_{\mu \nu}),
 \ee
 and
$\Phi_\mu$ is the basic vector field defined by \be \Phi_\mu
\stackrel{\rm def.}{=} {\gamma^\rho}_{\mu \rho}. \ee Here
$\alpha_1, \alpha_2,$ and $\alpha_3$ are constants determined by
M\o ller
 such that the theory coincides with general relativity in the weak fields:

\be \alpha_1=-{1 \over \kappa}, \qquad \alpha_2={\lambda \over
\kappa}, \qquad \alpha_3={1 \over \kappa}(1-\lambda), \ee where
$\kappa$ is the Einstein constant and  $\lambda$ is a free
dimensionless parameter\footnote{Throughout this paper we use the
relativistic units, $c=G=1$ and
 $\kappa=8\pi$.}. The same
choice of the parameters was also obtained by Hayashi and Nakano
\cite{HN}.

M\o ller applied the action principle to the Lagrangian density
(4) and obtained the field equation in the form \be G_{\mu \nu}
+H_{\mu \nu} = -{\kappa} T_{\mu \nu}, \ee \be F_{\mu \nu}=0, \ee
where the Einstein tensor $G_{\mu \nu}$ is defined by \be G_{\mu
\nu}=R_{\mu \nu}-{1 \over 2} g_{\mu \nu} R. \ee Here $H_{\mu \nu}$
and $F_{\mu \nu}$ are given by \be H_{\mu \nu} \stackrel{\rm
def.}{=} \lambda \left[ \gamma_{\rho \sigma \mu} {\gamma^{\rho
\sigma}}_\nu+\gamma_{\rho \sigma \mu} {\gamma_\nu}^{\rho
\sigma}+\gamma_{\rho \sigma \nu} {\gamma_\mu}^{\rho \sigma}+g_{\mu
\nu} \left( \gamma_{\rho \sigma \lambda} \gamma^{\lambda \sigma
\rho}-{1 \over 2} \gamma_{\rho \sigma \lambda} \gamma^{\rho \sigma
\lambda} \right) \right],
 \ee
and \be F_{\mu \nu} \stackrel{\rm def.}{=} \lambda \left[
\Phi_{\mu,\nu}-\Phi_{\nu,\mu} -\Phi_\rho \left({\gamma^\rho}_{\mu
\nu}-{\gamma^\rho}_{\nu \mu} \right)+ {{\gamma_{\mu
\nu}}^{\rho}}_{;\rho} \right], \ee and they are symmetric and skew
symmetric tensors, respectively.

M\o ller assumed that the energy-momentum tensor of matter fields
is symmetric. In the Hayashi-Nakano theory, however, the
energy-momentum tensor of spin-$1/2$ fundamental particles has
nonvanishing antisymmetric part arising from the effects due to
intrinsic spin, and the right-hand side of (9) does not vanish
when we take into account the possible effects of intrinsic spin.
Nevertheless, since in this paper we consider only solutions in
which the left hand side of (9) is identically vanishing, we refer
the tetrad theory of gravitation based on the choice of the
parameters, (7), as the Hayashi-Nakano-M\o ller (HNM) theory for
short.

It can be shown \cite{HS1} that the tensors, $H_{\mu \nu}$ and
 $F_{\mu \nu}$, consist of only those terms which are linear or quadratic
in the axial-vector part of the torsion tensor, $a_\mu$, defined
by \be a_\mu \stackrel{\rm def.}{=} {1 \over 3} \epsilon_{\mu \nu
\rho \sigma} \gamma^{\nu \rho \sigma}, \ee where $\epsilon_{\mu
\nu \rho \sigma}$ is defined by \be \epsilon_{\mu \nu \rho \sigma}
\stackrel{\rm def.}{=} (-g)^{1/2} \delta_{\mu \nu \rho \sigma} \ee
with $\delta_{\mu \nu \rho \sigma}$ being completely antisymmetric
and normalized as $\delta_{0123}=-1$. Therefore, both $H_{\mu
\nu}$ and $F_{\mu \nu}$ vanish if the $a_\mu$ is vanishing. In
other words, when the $a_\mu$ is found to vanish from the
antisymmetric part of the field equations, (9), the symmetric part
(8) coincides with the Einstein equation.

\newsection{Spherically symmetric nonsingular black hole solutions}

The tetrad space having three unknown functions of radial
coordinate with spherical symmetry   in spherical polar
coordinates, can be written as \cite{Ro}

 \be
\left(\il^\mu \right)= \left( \matrix{ iA & iDr & 0 & 0
\vspace{3mm} \cr 0 & B \sin\theta \cos\phi & \displaystyle{B \over
r}\cos\theta \cos\phi
 & -\displaystyle{B \sin\phi \over r \sin\theta} \vspace{3mm} \cr
0 & B \sin\theta \sin\phi & \displaystyle{B \over r}\cos\theta
\sin\phi
 & \displaystyle{B \cos\phi \over r \sin\theta} \vspace{3mm} \cr
0 & B \cos\theta & -\displaystyle{B \over r}\sin\theta  & 0 \cr }
\right), \ee where $i=\sqrt{-1}$.

Applying (15) to the field equations (8), (9) we note that the two
tensors $H_{\mu \nu}$ and  $F_{\mu \nu}$ are vanishing identically
regardless of the values of the functions $A$, $B$ and $D$. Thus
M\o ller field equations reduce for the tetrad (15) to Einstein's
equations. Then the field equations (8) and (9) take the form \ba
\kappa T_{0 0} \A= \A {1 \over r A^2 B^4}\Biggl[ \Biggl \{
\Biggl(3 D^2+8B'^2\Biggr) D-2\Biggl(2D B''+B'D'\Biggr)B \Biggr \}
r^3 B^2D- \nonu
\A \A \Biggl \{2\Biggl(D B''+B'D'\Biggr)B-5DB'^2\Biggr \} r^5
D^3-\Biggl(2BB''-3D^2-3B'^2 \Biggr) rB^4+\nonu
\A \A 2\Biggl(BD'-4DB'\Biggr)
r^4BD^3+2\Biggl(BD'-6DB'\Biggr)r^2B^3D-4B^5B' \Biggr], \nonu
\kappa T_{0 1} \A= \A  {D \over  A B^4} \Biggl[ \Biggl \{ 2 \Biggl
(D B''+B'D'\Biggr) B-5DB'^2  \Biggr \}
r^3D+\Biggl(2BB''-3D^2-3B'^2 \Biggr) rB^2-\nonu
\A \A  2\Biggl(B D'-4DB'\Biggr)r^2BD+4B^3B' \Biggr], \nonu
\kappa T_{1 1} \A= \A {1 \over r A B^4}\Biggl[ \Biggl \{ \Biggl(3
D^2+B'^2\Biggr) A+2BA'B' \Biggr \} r B^2- \Biggl \{ 2 \Biggl (D
B''+B'D'\Biggr) B-5DB'^2  \Biggr \} r^3AD + \nonu
\A \A  2\Biggl(B D'-4DB'\Biggr)r^2ABD-2AB^3B'-2B^4A' \Biggr],\nonu
\kappa T_{2 2} \A= \A {r \over  A^2 B^4} \Biggl[ \Biggl ( \Biggl
\{ \Biggl(D A''+3 A'D'\Biggr) B-3DA'B'  \Biggr \} A B D +\Biggl\{
\Biggl (2D B''+5B'D'\Biggr) B D- \nonu
\A \A \Biggl( D D''+D'^2 \Biggr) B^2-5D^2B'^2\Biggr \}
A^2-2B^2D^2A'^2 \Biggr) r^3+\nonu
\A \A \Biggl \{ \Biggl(B'^2 -3D^2\Biggr)
A^2-AB^2A''-B''BA^2+2B^2A'^2 \Biggr\}rB^2-\nonu
\A \A 2\Biggl \{ \Biggl( 3BD'-4DB'\Biggr)A-2BDA' \Biggr \}
r^2ABD+A^2B^3B'+AB^4A' \Biggr ], \nonu
T_{3 3} \A= \A sin \theta^2 T_{2 2}, \ea where
$A'=\displaystyle{dA \over dr}$, $B'=\displaystyle{dB \over dr}$
and  $D'=\displaystyle{dD \over dr}$.

Now we are going to find some special solutions to the partial
differential equations (16), assuming that the stress-energy
momentum tensor has the form \cite{Di} \ba
 {T_0}^0={T_1}^1, \nonu
  {T_2}^2={T_3}^3.
 \ea
  A first non-trivial solution
can be obtained by taking $D(r)=0$,
 and solving for $A(r)$ and $B(r)$, then we obtain
\newpage
\ba A \A=\A {1 \over \sqrt{1- \displaystyle{2m \over
R}\left(1-e^{-R^3/{r_1}^3}\right)}}, \nonu
B \A=\A \sqrt{1-\displaystyle{2m \over
R}\left(1-e^{-R^3/{r_1}^3}\right)}, \ea

where $R$ is a new radial coordinate  defined by $R=r/B$ and \ba
{r_1}^3 \A=\A r_g {r_0}^2,\nonu
 r_g\A=\A 2m,\nonu
{r_0}^2 \A=\A {3 \over 8\pi\epsilon_0}.
 \ea
 The form of
 the energy-momentum tensor is
  \ba
  {T_0}^0={T_1}^1 \A=\A
\epsilon_0 e^{-R^3/{r_1}^3},\nonu
{T_2}^2={T_3}^3 \A=\A \epsilon_0 e^{-R^3/{r_1}^3} \left(1-{3R^3
\over 2{r_1}^3} \right), \ea and the tetrad (15) takes the form
\be \left(\il^\mu \right)= \left( \matrix{ i \displaystyle{1 \over
\sqrt{1- \displaystyle{2m \over R}
\left(1-e^{-R^3/{r_1}^3}\right)}}  &0 & 0 & 0 \vspace{3mm} \cr 0 &
\sin\theta \cos\phi \sqrt{1-\displaystyle{2m \over
R}\left(1-e^{-R^3/{r_1}^3}\right)} & \displaystyle{\cos\theta
\cos\phi \over R}
 & -\displaystyle{ \sin\phi  \over R \sin\theta} \vspace{3mm} \cr
0 &  \sin\theta \sin\phi \sqrt{1-\displaystyle{2m \over
R}\left(1-e^{-R^3/{r_1}^3}\right)} & \displaystyle{\cos\theta
\sin\phi \over R}
 & \displaystyle{\cos\phi \over R \sin\theta} \vspace{3mm} \cr
0 &  \cos\theta \sqrt{1-\displaystyle{2m \over
R}\left(1-e^{-R^3/{r_1}^3}\right)} & -\displaystyle{\sin\theta
\over R} & 0 \cr } \right), \ee with the associated Riemannian
metric \be ds^2=-\eta_1 dt^2+{dR^2 \over \eta_1}+R^2 d\Omega^2,
\ee where \be \eta_1=\left[1-{2m \over
R}\left(1-e^{-R^3/r1^3}\right)\right], \ee and
${d\Omega^2=d\theta^2+\sin^2\theta d\phi^2}$, which is a static
spherically symmetric  nonsingular black hole solution \cite{Di}

A second non-trivial solution can be obtained by taking $A(r)=1$,
 $B(r)=1$, $D(r)\neq0$ and solving for $D(r)$. In this case the resulting field equations of
 (16) can be integrated directly to give
\be D(r)=\sqrt{{2m\over r3}\left(1-e^{-r^3/{r_1}^3}\right)}.
 \ee
Substituting for the value of $D(r)$ as given by (24) into (15),
we get
 \be
\left(\il^\mu \right)= \left( \matrix{ i &
i\sqrt{\displaystyle{2m\over r}\left(1-e^{-r^3/{r_1}^3}\right)} &
0 & 0 \vspace{3mm} \cr 0 &  \sin\theta \cos\phi &
\displaystyle{\cos\theta \cos\phi \over r}
 & -\displaystyle{ \sin\phi \over r \sin\theta} \vspace{3mm} \cr
0 &  \sin\theta \sin\phi & \displaystyle{\cos\theta \sin\phi \over
r}
 & \displaystyle{\cos\phi \over r \sin\theta} \vspace{3mm} \cr
0 &  \cos\theta & -\displaystyle{\sin\theta  \over r} & 0 \cr }
\right), \ee with the associated metric \be ds^2=-\left[1-{2m
\over r}\left(1-e^{-r^3/r1^3}\right)\right]dt^2-2\sqrt{{2m\over
r}\left(1-e^{-r^3/{r_1}^3}\right)}dr dt+dr^2+r^2 d\Omega^2,
 \ee
 it
is to be noted that m in the  metric (22) and (26) is a constant
of integration that will play the role of the mass producing the
field in the calculations of the energy. Also the form of the
stress-energy momentum tensor for this solution is the same as
given by (20).

Using the coordinate transformation \be dT=dt+{Dr \over 1-D^2r^2}
dr, \ee we can eliminate the cross term of (26) to obtain \be
ds^2=-\eta_2dT^2+{dr^2 \over \eta_2}+r^2 d\Omega^2,
 \ee
 where
$\eta_2$ is defined by (23) and $R=r$.

Thus we have two exact solutions of HNM field equations, each of
which leads to the same metric, a static spherically symmetric
nonsingular black hole in the spherical polar coordinate.

The solutions (18) and (24) are the exact solutions of the HNM
field equations. They  practically coincide with the Schwarzschild
solution for $r>>r1$ and  behave like the de Sitter solution, for
$r<<r1$.

As is clear from (20), the spherically symmetric stress-energy
momentum  tensor is really anisotropic. The difference between the
principle pressures \be {T_k}^k=-p_k, \ee correspond to the well
known anisotropic character of evolution of the space-time inside
a black hole undergoing a spherically symmetric gravitational
collapse \cite{ZN}. For $r<<r_1$ isotropization occurs and the
stress-energy momentum tensor takes the isotropic form \be
T_{\alpha \beta}=\epsilon g_{\alpha \beta}. \ee When $r\rightarrow
0$ the energy density tends to $\epsilon_0$. For $r>>r_1$ all the
components of the stress-energy momentum tensor tend to zero very
rapidly.

The important result obtained in this section is that we have been
able to derive two different solutions for HNM  theory; the
Riemannian metric associated with these two solutions are
identical, namely spherically symmetric nonsingular black hole.
Since HNM theory is a pure gravitational theory, the above two
solutions have to be equivalent in the sense that they describe
the same physical situation. In what follows we examine the
equivalence of these solutions by calculating the energy
associated with each of them, using the superpotential derived for
M\o ller's theory by Mikhail et al. \cite{MWHL}.
\newsection{The Energy Associated with each Solution}

The superpotential of the HNM theory is given by Mikhail et al.
\cite{MWHL} as \be {{\cal U}_\mu}^{\nu \lambda} ={(-g)^{1/2} \over
2 \kappa} {P_{\chi \rho \sigma}}^{\tau \nu \lambda}
\left[\Phi^\rho g^{\sigma \chi} g_{\mu \tau}
 -\lambda g_{\tau \mu} \gamma^{\chi \rho \sigma}
-(1-2 \lambda) g_{\tau \mu} \gamma^{\sigma \rho \chi}\right], \ee
where ${P_{\chi \rho \sigma}}^{\tau \nu \lambda}$ is \be {P_{\chi
\rho \sigma}}^{\tau \nu \lambda} \stackrel{\rm def.}{=}
{{\delta}_\chi}^\tau {g_{\rho \sigma}}^{\nu \lambda}+
{{\delta}_\rho}^\tau {g_{\sigma \chi}}^{\nu \lambda}-
{{\delta}_\sigma}^\tau {g_{\chi \rho}}^{\nu \lambda} \ee with
${g_{\rho \sigma}}^{\nu \lambda}$ being a tensor defined by \be
{g_{\rho \sigma}}^{\nu \lambda} \stackrel{\rm def.}{=}
{\delta_\rho}^\nu {\delta_\sigma}^\lambda- {\delta_\sigma}^\nu
{\delta_\rho}^\lambda. \ee The energy is expressed by the surface
integral \cite{Mo2} \be E=\lim_{r \rightarrow
\infty}\int_{r=constant} {{\cal U}_0}^{0 \alpha} n_\alpha dS, \ee
where $n_\alpha$ is the unit 3-vector normal to the surface
element ${\it dS}$.

Now we are in a position to calculate the energy associated with
 the two solutions (18) and (24) using the superpotential (31). As is
 clear from (34), the only components which contributes to the energy is ${{\cal U}_0}^{0
 \alpha}$. Thus substituting from the first solution (18) into
 (31) we obtain the following non-vanishing value
 \be
{{\cal U}_0}^{0 \alpha}={2X^\alpha \over \kappa r^2}{m \over
r}\left(1-e^{-r^3/r1^3}\right).
 \ee
 Substituting from (35) into
(34) we get \be E(r)=m \left(1-e^{-r^3/r1^3}\right). \ee This is a
satisfactory result and should be expected.

Now let us turn our attention to the second solution (24).
Calculating the necessary components of the superpotential, we get
\be {{\cal U}_0}^{0 \alpha}={4X^\alpha \over \kappa r^2}{m \over
r}\left(1-e^{-r^3/r1^3}\right). \ee Substituting from (37) into
(34) we get \be E(r)=2m \left(1-e^{-r^3/r1^3}\right). \ee That is
twice (36): it is clear from (36) and (38) that if $r\rightarrow
0$, $E(r)\rightarrow 0$ and as $r \rightarrow \infty$,
$E(r)\rightarrow m$ for (36) and $E(r)\rightarrow 2m$ for (38).
Note that $E(r)>0$ for all r:  $0\leq r < \infty$
\newsection{Main results and Discussion}

In this paper we have studied the nonsingular black hole
spherically symmetric solution in the HNM tetrad theory of
gravity. The axial vector part of the torsion, $a^\mu$ for these
solutions is identically vanishing.

Two different exact analytic solutions of the HNM field equations
are obtained for the case of spherical symmetry. The two solutions
give rise to the same Riemannian metric ( spherically symmetric
nonsingular black hole metric). The exact solutions (18) and (24)
represent a black hole which contain the de Sitter world instead
of a singularity. The stress-energy momentum tensor (20)
responsible for geometry describes a smooth transition from the
standard state at infinity to isotropic  state at $r\rightarrow 0$
through anisotropic state in intermediate region. This agrees with
the Poisson-Israel prediction concerning "non-inflationary
material at the interface" \cite{PI}.

It was shown by M\o ller \cite{Mo3} that a tetrad description of a
gravitational field equation allows a more satisfactory treatment
of the energy-momentum complex than does general relativity.
Therefore, we have applied the superpotential method given by
Mikhail et al.\cite{MWHL} to calculate the energy of the central
gravitating body. It is shown that the two solutions give two
different values
 of the energy content. The following suggestions may be
 considered to get out of this inconsistency:

 (a) The energy-momentum complex suggested by M\o ller \cite{Mo3} is
 not quite adequate; though it has the most satisfactory
 properties.\vspace{.3cm}\\
 (b) Many authors believe that a tetrad theory should describe
 more than a pure gravitation field. In fact; M\o ller himself
 \cite{Mo3} considered this possibility in his earlier trials to
 modify general relativity. In these theories, the most successful
 candidates for the description of the other physical phenomenon
 are the skew-symmetric tensors of the tetrad space, e.g.,
 $\Phi_{\mu;\nu}-\Phi_{\nu;\mu}$.  The most striking remark here
 is that: All the skew-symmetric tensors vanish for the first
 solution; but not all of them do so for the second one.  Some
 authors; e.g; \cite{Ei1,Mw}, believe that these tensors are related to
 the presence of an electromagnetic field. Others; e.g.; \cite{MN}
 believe  that these tensors are closely connected to the spin
 phenomenon. There are a lot of difficulties to claim that
 HNM theory deserves such a wider interpretation. This needs a
 lot of investigations before arriving at a concrete conclusion.\vspace{.3cm}\\
(c) Other possibility is that HNM theory is in need to be
generalized rather than to be reinterpreted. There are already
some generalizations of HNM theory. M\o ller himself considered
this possibility at the end of his paper \cite{Mo1}; by including
terms in the Lagrangian other than the simple quadratic terms.
S$\acute{a}$ez \cite{Se} has generalized HNM theory in a very
elegant and natural way into a scalar tetradic theories of
gravitation. In these theories the question is: Do the field
equations fix the tetradic geometry in the case of spherical
symmetry? This question was discussed in length by S$\acute{a}$ez
\cite{Se1}. The results of the present paper can be considered as
a first step to get a satisfactory answer to this question. Mayer
\cite{Me} has shown that M\o ller's theory is a special case of
the poincar$\acute{e}$ gauge theory constructed by Hehl et al.
\cite {HNV}. Thus poincar$\acute{e}$ gauge theory can be
considered as another satisfactory generalization of M\o ller's
theory.\vspace{.3cm}\\
 (d) Mikhail et al. \cite{MWHL} calculated
the energy of two spherically symmetric solutions and found that
the energy in one of the two solutions does not coincide with the
gravitational mass. Shirafuji et al. \cite{SNH} extended the
calculations to all the stationary asymptotically flat solutions
with spherical symmetry, dividing them into two classes, the one
in which the components
$({\al}_{\,{}^{{}^{{}^{\scriptstyle{0}}}}})$ and
$({\bl}_{\,{}^{{}^{{}^{\scriptstyle{\alpha}}}}})$ of the parallel
vector fields $({\al}_{\,{}^{{}^{{}^{\scriptstyle{\mu}}}}})$ tend
to zero faster than $1/\sqrt{r}$ for large r and the other, in
which  those components  go to zero as $1/\sqrt{r}$. It was found
that the equality of the energy and the gravitational mass holds
only in the first class. It is of interest to note that the two
tetrad structures  (21) and (25) have those properties, i.e., the
first tetrad structure (21)  the components
$({\al}_{\,{}^{{}^{{}^{\scriptstyle{0}}}}})$ and
$({\bl}_{\,{}^{{}^{{}^{\scriptstyle{\alpha}}}}})$ go to zero
faster than $1/\sqrt{r}$  for large r and so its energy is the
same as that obtained before \cite{Ra}. As for the second tetrad
structure (25) the components
$({\al}_{\,{}^{{}^{{}^{\scriptstyle{0}}}}})$ and
$({\bl}_{\,{}^{{}^{{}^{\scriptstyle{\alpha}}}}})$ go to zero as
$1/\sqrt{r}$. So its energy content is different from the energy
content of the first solution (18) and from that given by Yang
\cite{Yi} and  Radinschi \cite{Ra} by factor 2.

It is of interest to note that we have obtained  two exact
solutions for the partial differential equations (16) under some
special constraint. The general solution for the partial
differential equation in the case when the energy-momentum tensor
 does not vanish is not yet obtained. This will be studied in future
work.\\

\centerline{\Large{\bf Acknowledgements}}

The author would like to thank Professor I.F.I. Mikhail; Ain Shams
University,  Professor M.I. Wanas, Professor M. Melek; Cairo
University; Professor T. Shirafuji; Saitama University and
Professor K. Hayashi; Kitasato University for their stimulating
discussions .

\end{document}